\documentclass[preprint,showpacs,preprintnumbers,amsmath,amssymb,natbib,nofootinbib]{revtex4}
\usepackage{graphicx}% Include figure files
\usepackage{epsfig}
\usepackage{dcolumn}% Align table columns on decimal point
\usepackage{bm}% bold math

\def\be{\begin{equation}}
\def\ee{\end{equation}}

\def\ba{\begin{eqnarray}}
\def\ea{\end{eqnarray}}

\begin{document}
\title{Phase-space noncommutative formulation of Ozawa's uncertainty principle}

\author{Catarina Bastos\footnote{E-mail: catarina.bastos@ist.utl.pt}}
\affiliation{Instituto de Plasmas e Fus\~ao Nuclear, Instituto Superior T\'ecnico, Universidade de Lisboa, Avenida Rovisco Pais 1, 1049-001 Lisboa, Portugal}

\author{Alex E. Bernardini\footnote{ E-mail: alexeb@ufscar.br}}
\affiliation{Departamento de F\'isica, Universidade Federal de S\~ao Carlos, PO Box 676, 13565-905, S\~ao Carlos, SP, Brasil.}

\author{Orfeu Bertolami\footnote{Also at Centro de F\'isica do Porto, Rua do Campo Alegre 687, 4169-007 Porto, Portugal. E-mail: orfeu.bertolami@fc.up.pt}}
\affiliation{Departamento de F\'isica e Astronomia, Faculdade de Ci\^encias da Universidade do Porto, Rua do Campo Alegre, 687,4169-007 Porto, Portugal}

\author{{Nuno Costa Dias and Jo\~ao Nuno Prata}\footnote{Also at Grupo de F\'{\i}sica Matem\'atica, UL,
Avenida Prof. Gama Pinto 2, 1649-003, Lisboa, Portugal. E-mail: ncdias@meo.pt, joao.prata@mail.telepac.pt}}
\affiliation{Departamento de Matem\'{a}tica, Universidade Lus\'ofona de
Humanidades e Tecnologias Avenida Campo Grande, 376, 1749-024 Lisboa, Portugal}

\date{\today}

\begin{abstract}
%We generalize the Ozawa's measurement-disturbance relation for the phase-space noncommutative quantum mechanics. We show that in this case the measurement-disturbance relations have additional terms due to extra commutation relations between the variables. The noiseless case is not affected.
Ozawa's measurement-disturbance relation is generalized to a phase-space noncommutative extension of quantum mechanics.
It is shown that the measurement-disturbance relations have additional terms for backaction evading quadrature amplifiers and for noiseless quadrature transducers. Several distinctive features appear as a consequence of the noncommutative extension: measurement interactions which are noiseless, and observables which are undisturbed by a measurement, or of independent intervention in ordinary quantum mechanics, may acquire noise, become disturbed by the measurement, or no longer be an independent intervention in noncommutative quantum mechanics. It is also found that there can be states which violate Ozawa's universal noise-disturbance trade-off relation, but verify its noncommutative deformation.
\end{abstract}

\maketitle

1. {\it Introduction}:
Recently, there has been a great deal of discussion about the interpretation of the uncertainty principle \cite{Busch1,Busch2,Ozawa1} and its possible experimental violation \cite{Rozema,Hasegawa,Ozawa0}. On the one hand, in his well-known $\gamma$-ray thought experiment \cite{Heisenberg1}, Heisenberg relates the accuracy of an appropriate position measurement to the disturbance of the particle's momentum, for a system in a state $\psi$, so that
\begin{equation}
\epsilon (\widehat{X}, \psi) \chi (\widehat{P}, \psi) \ge {{|\langle \psi | ~ \left[\widehat{X}, \widehat{P} \right] ~ | \psi\rangle|}\over 2}~,
\label{eq0}
\end{equation}
where $\epsilon (\widehat{X}, \psi)$ is the noise of the $\widehat{X}$ measurement and $\chi (\widehat{P}, \psi)$ is the disturbance on $\widehat{P}$ due to that measurement. Rather recently, Busch {\em et al.} \cite{Busch1} claim to have proved a rigorous version of Heisenberg's uncertainty principle considering error and disturbance measurements as state independent quantities, which contrasts with the state dependent preparation uncertainty principles. On the other hand, state dependent quantities have also been considered \cite{Korzekwa,Ozawa,Ozawa4,Cyril1}, such as Ozawa's noise-disturbance trade-off relation \cite{Ozawa4}. He considers a system composed by an object and a measuring device, the probe, initially prepared as $\Psi = \psi \otimes \xi$, where $\psi$ and $\xi$ describe the object and the probe, respectively. Working in the Heisenberg picture, Ozawa then introduces the noise operator, $\widehat{N}(\widehat{A})$, and the disturbance operator, $\widehat{D}(\widehat{B})$, related to the observables $A$ and $B$, respectively. These are self-adjoint operators, defined as
\begin{equation}
\widehat{N}(A) = \widehat{M}^{out} - \widehat{A}^{in}\hspace{0.2cm}, \hspace{0.2cm} \widehat{D}(B) = \widehat{B}^{out} - \widehat{B}^{in}~.
\label{eq1}
\end{equation}
Here $\widehat{A}^{in}=\widehat{A} \otimes \widehat{I}, \widehat{B}^{in}=\widehat{B} \otimes \widehat{I}$ are observables $\widehat{A}$ and $\widehat{B}$ prior to the measurement interaction, $\widehat{B}^{out} =\widehat{U}^{\dagger} (\widehat{B} \otimes \widehat{I}) \widehat{U}$ is the observable $\widehat{B}$ immediately after the measurement, and $\widehat{M}$ is the probe observable, that is, the measurement corresponding to the observable $\widehat{A}$. $\widehat{U}$ is an unitary time evolution operator that acts during the measuring interaction. Clearly, $\widehat{M}^{in} = \widehat{I} \otimes \widehat{M}$ and $\widehat{M}^{out} = \widehat{U}^{\dagger} (\widehat{I} \otimes \widehat{M}) \widehat{U}$. For more details about the measurement interaction see Ref.~\cite{Ozawa4}. The noise $\epsilon (\widehat{A}, \psi)$ and disturbance $\chi (\widehat{B}, \psi)$ are defined by \cite{Ozawa4}:
\begin{equation}
\epsilon (\widehat{A}, \psi)^2 = \langle \Psi | \widehat{N}(\widehat{A})^2 | \Psi\rangle\hspace{0.2cm} , \hspace{0.2cm} \chi (\widehat{B}, \psi)^2 =  \langle\Psi | \widehat{D}(\widehat{B})^2 | \Psi\rangle~.
\label{eq2}
\end{equation}
Since $\widehat{M}$ and $\widehat{B}$ are observables in different systems, they commute. Thus, using $\left[\widehat{M}^{out}, \widehat{B}^{out} \right]=0$, Eq. (\ref{eq2}), the triangle and the Cauchy-Schwartz inequalities, one obtains Ozawa's uncertainty principle (OUP),
\begin{equation}
\epsilon (\widehat{A}, \psi) \chi (\widehat{B}, \psi) + {1\over2}{\left| \langle \left[\widehat{N}(\widehat{A}),\widehat{B}^{in} \right]\rangle + \langle \left[\widehat{A}^{in} , \widehat{D}(\widehat{B}) \right]\rangle\right|}\ge {1\over2}{| \langle\psi| ~\left[\widehat{A}, \widehat{B} \right] ~| \psi\rangle}~,
\label{eq3}
\end{equation}
where one has used, as suggested in Ref. \cite{Ozawa4}, the notation, $\langle \widehat{C}\rangle$, to denote $\langle\Psi | \widehat{C} | \Psi\rangle$.
If $\left| \langle \left[\widehat{N}(\widehat{A}),\widehat{B}^{in} \right]\rangle + \langle \left[\widehat{A}^{in} , \widehat{D}(\widehat{B}) \right]\rangle\right|=0$, then the Heisenberg noise-disturbance uncertainty relation, Eq. (\ref{eq0}), holds. Ozawa defined such a measuring interaction to be of {\it independent intervention} for the pair $(\widehat{A},\widehat{B})$.

From the triangle and the Cauchy-Schwartz inequalities one has:
\begin{equation}
\left| \langle \left[\widehat{N}(\widehat{A}),\widehat{B}^{in} \right]\rangle + \langle \left[\widehat{A}^{in} , \widehat{D}(\widehat{B}) \right]\rangle\right|  \le 2 \epsilon (\widehat{A}, \psi) \sigma (\widehat{B}, \psi) + 2 \sigma (\widehat{A}, \psi) \chi (\widehat{B}, \psi)~,
\label{eq4}
\end{equation}
and upon substitution of Eq.~(\ref{eq2}) into Eq.~(\ref{eq3}) one obtains:
\begin{equation}
\epsilon (\widehat{A}, \psi) \chi (\widehat{B}, \psi) +\epsilon (\widehat{A}, \psi) \sigma (\widehat{B},\psi) +  \sigma (\widehat{A}, \psi) \chi (\widehat{B}, \psi) \ge {{| \langle\psi|
~\left[\widehat{A}, \widehat{B} \right] ~| \psi\rangle |}\over2}~. \label{eq5}
\end{equation}
Here $\sigma (\widehat{C}, \psi)$ denotes the mean square deviation for observable $C$: $\sigma (\widehat{C}, \psi)= \langle \psi| \left(\widehat{C} - \langle \widehat{C} \rangle \right)^2 | \psi \rangle^{1/2}$. Experimentally,  weak measurements \cite{Rozema} and $3-$state mode systems \cite{Hasegawa, Cyril2}, indicate that violations of inequality Eq. (\ref{eq3}) are found and Eq.~(\ref{eq5}) is shown to be a more accurate description of the experimental data.

Hereon we investigate whether noncommutative extensions of quantum
mechanics (NCQM) yield corrections to the OUP. NCQM corresponds to
a non-relativistic one-particle sector of NC quantum field
theories \cite{Douglas}, which emerge in the context of string
theory and quantum gravity \cite{Connes}. Actually, the idea that
space-time has noncommutativity features was suggested long ago as
a way to regularize quantum field theories
\cite{Snyder,Heisenberg,Yang}. However, this possibility was
disconsidered for a while due to the development of
renormalization techniques and to certain undesirable features of
NC theories, such as the breakdown of Lorentz invariance
\cite{Carrol,Bertolami-Guisado}. More recently, noncommutativity
was revived in the context of quantization of gravity
\cite{Freidel}. The discovery that the low-energy effective theory
of a D-brane in the background of a Neveu-Schwarz B field lives on
a space with spatial noncommutativity  has triggered further the
interest in this putative feature of space-time
\cite{Douglas,Douglas2,Connes}. From another perspective, a simple
heuristic argument, based on Heisenberg's uncertainty principle,
the equivalence principle and the Schwarzschild metric, shows that
the Planck length seems to be a lower bound on the precision of a
position measurement \cite{Rosenbaum}. This reenforces the point
of view that a new NC geometry of space-time may emerge at a
fundamental level \cite{Douglas,Connes2,Martinetti,Szabo}.

From another perspective, NC deformations of the Heisenberg-Weyl
(HW) algebra have been investigated in the context of quantum
cosmology and are shown to have relevant implications on the
thermodynamic stability of black holes and as a possible
regularization of the black hole singularities \cite{Bastos3}. In
the context of QM, phase-space noncommutativity could induce
violations of Robertson-Schr\"odinger uncertainty principle
\cite{Bastos4} and also work as a source of Gaussian entanglement
\cite{Bastos5}. Actually, a matrix and Robertson-Schr\"odinger
version of the OUP has been discussed in Ref. \cite{Bastos6}.

A deformation of the HW algebra may also appear as a consequence of quantizing systems with constraints (see e.g. \cite{Nakamura}).

Phase-space NCQM follows from the modified HW algebra,
\be\label{NCQM}
[\widehat{X},\widehat{Y}]=i\theta\hspace{1cm},\hspace{1cm} [\widehat{P}_{X},\widehat{P}_Y]=i\eta\hspace{1cm},\hspace{1cm} [\widehat{X},\widehat{P}_X]=[\widehat{Y},\widehat{P}_Y]=i\hbar~,
\ee
where the NC parameters $\theta$ and $\eta$ are real constants \cite{Bastos1}.
%For more details about this canonical extension of QM see Ref.~\cite{Bastos1}

The paper is organized as follows. In order to implement the NC corrections to the OUP predictions, we start by setting our notation and describing the so-called backaction evading (BAE) quadrature amplifiers in the next section. In section 3, we address the noncommutative deformation of the BAE interaction. In section 4, similar calculations are performed for another type of measurement interaction - the noiseless quadrature transducers. Finally, in section 5, we state our conclusions.

2. {\it Backaction evading quadrature amplifiers}: In the sequel, Latin indices $i,j,k, \cdots$ take values in the set $\left\{1, \cdots,n \right\}$, whereas Greek indices $\alpha, \beta, \cdots$ run from $1$ to $2n$. Let $\widehat{A}_i^{in}$ and $\widehat{B}_j^{in}$, $i,j=1, \cdots, n$ denote a set of self-adjoint operators such that
\begin{equation}
\left[\widehat{A}_i^{in},\widehat{A}_j^{in} \right] = \left[\widehat{B}_i^{in},\widehat{B}_j^{in} \right] =0, \hspace{1 cm} \left[\widehat{A}_i^{in},\widehat{B}_j^{in} \right] = i C_{ij},
\label{eq5B}
\end{equation}
for $i,j=1, \cdots, n$, and where $C=\left\{C_{ij} \right\}$ is a real non-vanishing matrix.
One may write the operators collectively as
\begin{equation}
\widehat{Z}^{in} = \left(\widehat{A}_1^{in}, \cdots, \widehat{A}_n^{in},\widehat{B}_1^{in}, \cdots, \widehat{B}_n^{in} \right),
\label{eq6}
\end{equation}
satisfying the commutation relations
\begin{equation}
\left[\widehat{Z}_{\alpha}^{in},\widehat{Z}_{\beta}^{in} \right] = i G_{\alpha \beta}, \hspace{1 cm} \alpha, \beta =1, \cdots, 2n,
\label{eq7}
\end{equation}
with $G=\left\{G_{\alpha \beta}\right\}$ the skew-symmetric matrix:
\begin{equation}
G= \left(
\begin{array}{c c}
0 & C\\
- C^T & 0
\end{array}
\right).
\label{eq7.1}
\end{equation}
If $\widehat{A}^{in}$ denotes the object's position, $\widehat{X}^{in}$, and, $\widehat{B}^{in}$, its momentum, $\widehat{P}^{in}$, then $\left\{C_{ij} \right\}$ becomes the identity matrix $\left\{\delta_{ij} \right\}$ and $\left\{G_{\alpha \beta} \right\}$ becomes the standard symplectic matrix $J=\left\{J_{\alpha \beta} \right\}$:
\begin{equation}
J= \left(
\begin{array}{c c}
0 & I\\
- I & 0
\end{array}
\right).
\label{eq7.2}
\end{equation}
Let $\widehat{M}^{out} = \left(\widehat{M}^{out}_1, \cdots, \widehat{M}^{out}_n \right)$ denote the outputs of the probe observable. The noise operators are defined by
\begin{equation}
\widehat{N}_i= \widehat{N}_i (\widehat{A}_i)= \widehat{M}_i^{out} - \widehat{A}_i^{in}, \hspace{1 cm} i=1, \cdots, n.
\label{eq7.3}
\end{equation}
Similarly, the disturbance on the observable $\widehat{B}$ is given by:
\begin{equation}
\widehat{D}_i= \widehat{D}_i (\widehat{B}_i)= \widehat{B}_i^{out} - \widehat{B}_i^{in}, \hspace{1 cm} i=1, \cdots, n.
\label{eq7.4}
\end{equation}
We can write these quantities collectively as
\begin{equation}
\widehat{K} = \left(\widehat{N}_1, \cdots, \widehat{N}_n, \widehat{D}_1, \cdots, \widehat{D}_n \right).
\label{eq10}
\end{equation}
If we write
\begin{equation}
\widehat{Z}^{out} = \left(\widehat{M}^{out}_1, \cdots, \widehat{M}^{out}_n,\widehat{B}_1^{out}, \cdots, \widehat{B}_n^{out} \right),
\label{eq13}
\end{equation}
then according to Ozawa \cite{Ozawa,Ozawa4} one has:
\begin{equation}
\left[\widehat{Z}_{\alpha}^{out}, \widehat{Z}_{\beta}^{out} \right]=0, \hspace{1 cm} \alpha , \beta =1, \cdots, 2n,
\label{eq14}
\end{equation}
and
\begin{equation}
\widehat{Z}^{out}= \widehat{Z}^{in} + \widehat{K}.
\label{eq15}
\end{equation}

To study the validity of Eq.~(\ref{eq5}), Ozawa \cite{Ozawa4} considered the system quadrature operators $\widehat{X}_a,\widehat{P}_{X_a}$ and the probe operators $\widehat{X}_b,\widehat{P}_{X_b}$ obeying to the commutation relations
\begin{equation}
\left[\widehat{X}_a,\widehat{P}_{X_a} \right]= \left[\widehat{X}_b,\widehat{P}_{X_b} \right] = i\hbar~.
\label{eq16}
\end{equation}
For the sake of generality, we will keep the Planck constant $\hbar$ arbitrary here. Ozawa \cite{Ozawa4} considered $\hbar =1/2$. The measuring interaction is given by
\begin{equation}
\left\{
\begin{array}{l}
\widehat{X}_a^{out}= \widehat{X}_a^{in}\\
\widehat{X}_b^{out} = \widehat{X}_b^{in} +  G \widehat{X}_a^{in}\\
\widehat{P}_{X_a}^{out} = \widehat{P}_{X_a}^{in} - G \widehat{P}_{X_b}^{in}\\
\widehat{P}_{X_b}^{out} = \widehat{P}_{X_b}^{in}~,
\end{array}
\right.
\label{eq17}
\end{equation}
where the factor $G$ is the gain associated to the measurement. The probe observable is then set to $\widehat{M}= {1\over G} \widehat{X}_b$, and thus
\begin{equation}
\widehat{M}^{out} = \widehat{X}_a^{in} + {1\over G} \widehat{X}_b^{in}.
\label{eq18}
\end{equation}
Moreover, the noise and disturbance are given by
\begin{eqnarray}
\widehat{N} (X_a) &=&{1\over G} \widehat{X}_b^{in},\nonumber\\
\widehat{D} (X_a) &=&0,\nonumber\\
\widehat{D} (P_{X_a}) &=& - G \widehat{P}_{X_b}^{in}.
\label{eq19}
\end{eqnarray}

This measuring model is referred to as backaction evading quadrature (BAE) amplifier and is considered here for the measurement of the quadrature operator $X_a$ in a $1-$dimensional case.

In order to implement the NC commutation relations, we must first consider a $2-$dimensional extension of the BAE model. We define the additional degrees of freedom $\widehat{Y}_a$ and $\widehat{Y}_b$ for the system and the probe and the conjugate variables $\widehat{P}_{Y_a}$ and $\widehat{P}_{Y_b}$, obeying the same commutation relations (\ref{eq16}):
\begin{equation}
\left[\widehat{X}_a,\widehat{P}_{X_a} \right]=\left[\widehat{Y}_a,\widehat{P}_{Y_a} \right] =\left[\widehat{X}_b,\widehat{P}_{X_b} \right]=\left[\widehat{Y}_b,\widehat{P}_{Y_b} \right]= i \hbar,
\label{eqcomments2}
\end{equation}
while all the remaining commutators vanish.  Let $\widehat{H}$ be the Hamiltonian operator
\begin{equation}
\widehat{H} = \alpha \left(\widehat{P}_{X_b} \widehat{X}_a +\widehat{P}_{Y_b} \widehat{Y}_a \right),
\label{eqcomments1}
\end{equation}
where $\alpha$ is some constant with dimensions $(time)^{-1}$. It generates the unitary transformation
\begin{equation}
\widehat{U} (t) = e^{\frac{it}{\hbar} \widehat{H}}
\label{eqcomments1.1}
\end{equation}
which models the measurement interaction during a time interval $t \in \left[0, T \right]$.

In view of the commutation relations (\ref{eqcomments2}), there are no ordering ambiguities in the Hamiltonian, Eq. (\ref{eqcomments1}).

The equations of motion are
\begin{equation}
{{d \widehat{Z}}\over dt} = {1\over{i \hbar}} \left[\widehat{Z},\widehat{H} \right]
\label{eqcomments3}
\end{equation}
for the observable $\widehat{Z}$. From Eqs. (\ref{eqcomments1})-(\ref{eqcomments3}) one obtains:
\begin{equation}
\left\{
\begin{array}{l}
{{d\widehat{X}_a}\over dt}= 0\\
{{d \widehat{Y}_a}\over dt} = 0\\
{{d \widehat{X}_b}\over dt} = \alpha \widehat{X}_a\\
{{d \widehat{Y}_b}\over dt} = \alpha \widehat{Y}_a\\
{{d \widehat{P}_{X_a}}\over dt} = -\alpha \widehat{P}_{X_b}\\
{{d \widehat{P}_{Y_a}}\over dt} = -\alpha \widehat{P}_{Y_b}\\
{{d \widehat{P}_{X_b}}\over dt} =0\\
{{d \widehat{P}_{Y_b}}\over dt} =0~.
\end{array}
\right.
\label{eqcomments4}
\end{equation}
At time $t=0$, just before the measurement interaction is switched on, one has $ \widehat{X}_a (0)= \widehat{X}_a^{in}$, $ \widehat{Y}_a (0)= \widehat{Y}_a^{in}$, $ \widehat{X}_b (0)= \widehat{X}_b^{in}$, etc.

The solution of Eqs. (\ref{eqcomments4}) is then:
\begin{equation}
\left\{
\begin{array}{l}
\widehat{X}_a (t) = \widehat{X}_a^{in}\\
\widehat{Y}_a (t) = \widehat{Y}_a^{in}\\
\widehat{X}_b (t) = \widehat{X}_b^{in} + t\alpha \widehat{X}_a^{in} \\
\widehat{Y}_b (t) = \widehat{Y}_b^{in} + t\alpha\widehat{Y}_a^{in} \\
\widehat{P}_{X_a} (t) = \widehat{P}_{X_a}^{in} - t \alpha\widehat{P}_{X_b}^{in} \\
\widehat{P}_{Y_a} (t) = \widehat{P}_{Y_a}^{in} - t \alpha \widehat{P}_{Y_b}^{in} \\
\widehat{P}_{X_b} (t) = \widehat{P}_{X_b}^{in}  \\
\widehat{P}_{Y_b} (t) = \widehat{P}_{Y_b}^{in}~.
\end{array}
\right.
\label{eqcomments5}
\end{equation}
Let $T$ be the infinitesimal duration of the interaction. Thus, one sets $ \widehat{X}_a (T)= \widehat{X}_a^{out}$, $ \widehat{Y}_a (T)= \widehat{Y}_a^{out}$, $ \widehat{X}_b (T)= \widehat{X}_b^{out}$, etc. Let also $G= \alpha T$ be a dimensionless constant. Hence:
\begin{equation}
\left\{
\begin{array}{l}
\widehat{X}_a^{out} = \widehat{X}_a^{in}\\
\widehat{Y}_a^{out} = \widehat{Y}_a^{in}\\
\widehat{X}_b^{out} = \widehat{X}_b^{in} + G\widehat{X}_a^{in} \\
\widehat{Y}_b^{out} = \widehat{Y}_b^{in} + G\widehat{Y}_a^{in} \\
\widehat{P}_{X_a}^{out} = \widehat{P}_{X_a}^{in} - G\widehat{P}_{X_b}^{in} \\
\widehat{P}_{Y_a}^{out} = \widehat{P}_{Y_a}^{in} - G\widehat{P}_{Y_b}^{in} \\
\widehat{P}_{X_b}^{out} = \widehat{P}_{X_b}^{in}  \\
\widehat{P}_{Y_b}^{out} = \widehat{P}_{Y_b}^{in}~,
\end{array}
\right.
\label{eqcomments6}
\end{equation}
which is the BAE interaction for the $2$-dimensional case.

The probe observables are now set to
\be\label{eq21}
\widehat{M}=\left({\widehat{X}_b\over G}, {\widehat{Y}_b\over G}\right)~,
\ee
and then the vector $\widehat{K}$, which is the generalized vector describing the noise and the disturbance of the measurement, is given by
\be\label{eq22}
\widehat{K}=\left({1\over G}\widehat{X}_b, {1\over G}\widehat{Y}_b, -G \widehat{P}_{x_b}, -G \widehat{P}_{y_b}\right)~.
\ee

3. {\it The noncommutative extension of Ozawa's uncertainty principle}:
We now turn to the noncommutative algebra
\ba
\left[\widehat{X}_a,\widehat{Y}_a \right] = \left[\widehat{X}_b,\widehat{Y}_b \right] =i \theta~, \label{eq23.1}\\
\left[\widehat{P}_{X_a},\widehat{P}_{Y_a} \right] = \left[\widehat{P}_{X_b},\widehat{P}_{Y_b} \right] =i \eta~, \label{eq23.2}\\
\left[\widehat{X}_a,\widehat{P}_{X_a} \right] = \left[\widehat{Y}_a,\widehat{P}_{Y_a} \right] =i \hbar~, \label{eq23.3}\\
\left[\widehat{X}_b,\widehat{P}_{X_b} \right] = \left[\widehat{Y}_b,\widehat{P}_{Y_b} \right] =i \hbar~, \label{eq23.4}
\ea
and all remaining commutators vanish.

It is reasonable to anticipate that, if one performs a measurement of, say, $\widehat{X}_a$, then:

(a) There may be an associated noise, depending on the nature of measurement interaction, e.g. BAE or noiseless;

(b) One should expect a disturbance on the canonically conjugate momentum, $\widehat{P}_{X_a}$, due to HW algebra, and consequently on $\widehat{P}_{Y_a}$, since in the NC framework momenta do not commute, i. e. $[\widehat{P}_X,\widehat{P}_Y]=i\eta$;

(c) Moreover, one has an additional disturbance on $\widehat{Y}_a$ due to the NC relation $[\widehat{X},\widehat{Y}]=i\theta$, and indirectly on $\widehat{P}_{Y_a}$, due to the standard commutation relation $[\widehat{Y}, \widehat{P}_Y]=i\hbar$.

 We assume the same Hamiltonian, Eq. (\ref{eqcomments1}), as above. Notice that there are still no ordering ambiguities in the Hamiltonian.

The equations of motion then become:
\begin{equation}
\left\{
\begin{array}{l}
{{d \widehat{X}_a}\over dt} = {{\alpha \theta}\over\hbar} \widehat{P}_{Y_b}\\
{{d \widehat{Y}_a}\over dt} = - {{\alpha \theta}\over\hbar} \widehat{P}_{X_b}\\
{{d \widehat{X}_b}\over dt} = \alpha  \widehat{X}_a \\
{{d \widehat{Y}_b}\over dt} = \alpha  \widehat{Y}_a \\
{{d \widehat{P}_{X_a}}\over dt} = - \alpha \widehat{P}_{X_b} \\
{{d \widehat{P}_{Y_a}}\over dt} = - \alpha \widehat{P}_{Y_b} \\
{{d \widehat{P}_{X_b}}\over dt} =  {{\alpha \eta}\over\hbar} \widehat{Y}_a \\
{{d \widehat{P}_{Y_b}}\over dt} =  - {{\alpha \eta}\over\hbar} \widehat{X}_a~.
\end{array}
\right.
\label{eqcomments8}
\end{equation}
Again setting $ \widehat{X}_a (0)= \widehat{X}_a^{in}$, $ \widehat{Y}_a (0)= \widehat{Y}_a^{in}$, $ \widehat{X}_b (0)= \widehat{X}_b^{in}$, etc, the solution to the previous system of equations reads:
\begin{equation}
\left\{
\begin{array}{l}
\widehat{X}_a (t) = \widehat{X}_a^{in} \cos \left( \frac{\alpha t \sqrt{\theta \eta}}{\hbar}\right) + \sqrt{\frac{\theta}{\eta}} \widehat{P}_{Y_b}^{in} \sin  \left( \frac{\alpha t \sqrt{\theta \eta}}{\hbar}\right)\\
\widehat{Y}_a (t) = \widehat{Y}_a^{in} \cos \left( \frac{\alpha t \sqrt{\theta \eta}}{\hbar}\right) - \sqrt{\frac{\theta}{\eta}} \widehat{P}_{X_b}^{in} \sin  \left( \frac{\alpha t \sqrt{\theta \eta}}{\hbar}\right)\\
\widehat{X}_b (t) = \widehat{X}_b^{in}+ \frac{\hbar}{\sqrt{\theta \eta}} \widehat{X}_a^{in} \sin \left( \frac{\alpha t \sqrt{\theta \eta}}{\hbar}\right) + \frac{2 \hbar}{\eta} \widehat{P}_{Y_b}^{in} \sin^2  \left( \frac{\alpha t \sqrt{\theta \eta}}{2 \hbar}\right)\\
\widehat{Y}_b (t) = \widehat{Y}_b^{in}+ \frac{\hbar}{\sqrt{\theta \eta}} \widehat{Y}_a^{in} \sin \left( \frac{\alpha t \sqrt{\theta \eta}}{\hbar}\right) - \frac{2 \hbar}{\eta} \widehat{P}_{X_b}^{in} \sin^2  \left( \frac{\alpha t \sqrt{\theta \eta}}{2 \hbar}\right)\\
\widehat{P}_{X_a} (t) = \widehat{P}_{X_a}^{in}- \frac{\hbar}{\sqrt{\theta \eta}} \widehat{P}_{X_b}^{in} \sin \left( \frac{\alpha t \sqrt{\theta \eta}}{\hbar}\right) - \frac{2 \hbar}{\theta} \widehat{Y}_a^{in} \sin^2  \left( \frac{\alpha t \sqrt{\theta \eta}}{2 \hbar}\right)\\
\widehat{P}_{Y_a} (t) = \widehat{P}_{Y_a}^{in}- \frac{\hbar}{\sqrt{\theta \eta}} \widehat{P}_{Y_b}^{in} \sin \left( \frac{\alpha t \sqrt{\theta \eta}}{\hbar}\right) + \frac{2 \hbar}{\theta} \widehat{X}_a^{in} \sin^2  \left( \frac{\alpha t \sqrt{\theta \eta}}{2 \hbar}\right)\\
\widehat{P}_{X_b} (t) =  \widehat{P}_{X_b}^{in} \cos \left( \frac{\alpha t \sqrt{\theta \eta}}{\hbar}\right) + \sqrt{ \frac{\eta}{\theta}} \widehat{Y}_a^{in} \sin  \left( \frac{\alpha t \sqrt{\theta \eta}}{\hbar}\right)\\
\widehat{P}_{Y_b} (t) =  \widehat{P}_{Y_b}^{in} \cos \left( \frac{\alpha t \sqrt{\theta \eta}}{\hbar}\right) - \sqrt{ \frac{\eta}{\theta}} \widehat{X}_a^{in} \sin  \left( \frac{\alpha t \sqrt{\theta \eta}}{\hbar}\right)~.
\end{array}
\right.
\label{eqcomments8}
\end{equation}
As previously, $ \widehat{X}_a (T)= \widehat{X}_a^{out}$, $ \widehat{Y}_a (T)= \widehat{Y}_a^{out}$, $ \widehat{X}_b (T)= \widehat{X}_b^{out}$, etc. We thus obtain:
\begin{equation}
\left\{
\begin{array}{l}
\widehat{X}_a^{out} = \widehat{X}_a^{in} \cos \left( \frac{G \sqrt{\theta \eta}}{\hbar}\right) + \sqrt{\frac{\theta}{\eta}} \widehat{P}_{Y_b}^{in} \sin  \left( \frac{G \sqrt{\theta \eta}}{\hbar}\right)\\
\widehat{Y}_a^{out} = \widehat{Y}_a^{in} \cos \left( \frac{G \sqrt{\theta \eta}}{\hbar}\right) - \sqrt{\frac{\theta}{\eta}} \widehat{P}_{X_b}^{in} \sin  \left( \frac{G \sqrt{\theta \eta}}{\hbar}\right)\\
\widehat{X}_b^{out} = \widehat{X}_b^{in}+ \frac{\hbar}{\sqrt{\theta \eta}} \widehat{X}_a^{in} \sin \left( \frac{G \sqrt{\theta \eta}}{\hbar}\right) + \frac{2 \hbar}{\eta} \widehat{P}_{Y_b}^{in} \sin^2  \left( \frac{G \sqrt{\theta \eta}}{2 \hbar}\right)\\
\widehat{Y}_b^{out}= \widehat{Y}_b^{in}+ \frac{\hbar}{\sqrt{\theta \eta}} \widehat{Y}_a^{in} \sin \left( \frac{G \sqrt{\theta \eta}}{\hbar}\right) - \frac{2 \hbar}{\eta} \widehat{P}_{X_b}^{in} \sin^2  \left( \frac{G \sqrt{\theta \eta}}{2 \hbar}\right)\\
\widehat{P}_{X_a}^{out} = \widehat{P}_{X_a}^{in}- \frac{\hbar}{\sqrt{\theta \eta}} \widehat{P}_{X_b}^{in} \sin \left( \frac{G \sqrt{\theta \eta}}{\hbar}\right) - \frac{2 \hbar}{\theta} \widehat{Y}_a^{in} \sin^2  \left( \frac{G \sqrt{\theta \eta}}{2 \hbar}\right)\\
\widehat{P}_{Y_a}^{out} = \widehat{P}_{Y_a}^{in}- \frac{\hbar}{\sqrt{\theta \eta}} \widehat{P}_{Y_b}^{in} \sin \left( \frac{G \sqrt{\theta \eta}}{\hbar}\right) + \frac{2 \hbar}{\theta} \widehat{X}_a^{in} \sin^2  \left( \frac{G \sqrt{\theta \eta}}{2 \hbar}\right)\\
\widehat{P}_{X_b}^{out} =  \widehat{P}_{X_b}^{in} \cos \left( \frac{G \sqrt{\theta \eta}}{\hbar}\right) + \sqrt{ \frac{\eta}{\theta}} \widehat{Y}_a^{in} \sin  \left( \frac{G \sqrt{\theta \eta}}{\hbar}\right)\\
\widehat{P}_{Y_b}^{out} =  \widehat{P}_{Y_b}^{in} \cos \left( \frac{G \sqrt{\theta \eta}}{\hbar}\right) - \sqrt{ \frac{\eta}{\theta}} \widehat{X}_a^{in} \sin  \left( \frac{G \sqrt{\theta \eta}}{\hbar}\right)~.
\end{array}
\right.
\label{eqcomments9}
\end{equation}

Since the duration of the measurement interaction is infinitesimal ($G<<1$) and the noncommutative parameters are presumably small ($\frac{\sqrt{\theta \eta}}{\hbar} <<1$) \cite{Bertolami}, one keeps only the lowest order terms in the previous expressions to get:
\begin{equation}
\left\{
\begin{array}{l}
\widehat{X}_a^{out} \sim \widehat{X}_a^{in} + {{G \theta}\over\hbar} \widehat{P}_{Y_b}^{in}\\
\widehat{Y}_a^{out} \sim \widehat{Y}_a^{in}  - {{G \theta}\over\hbar} \widehat{P}_{X_b}^{in} \\
\widehat{X}_b^{out} \sim \widehat{X}_b^{in}+ G \widehat{X}_a^{in}  + {{\theta G^2}\over 2 \hbar} \widehat{P}_{Y_b}^{in} \\
\widehat{Y}_b^{out}\sim \widehat{Y}_b^{in}+ G \widehat{Y}_a^{in}  - {{\theta G^2}\over 2 \hbar} \widehat{P}_{X_b}^{in} \\
\widehat{P}_{X_a}^{out} \sim \widehat{P}_{X_a}^{in}- G  \widehat{P}_{X_b}^{in}  - {{\eta G^2}\over2 \hbar} \widehat{Y}_a^{in}\\
\widehat{P}_{Y_a}^{out} \sim \widehat{P}_{Y_a}^{in}- G \widehat{P}_{Y_b}^{in}  + {{\eta G^2}\over2 \hbar} \widehat{X}_a^{in} \\
\widehat{P}_{X_b}^{out} \sim  \widehat{P}_{X_b}^{in} +  {{G \eta}\over\hbar} \widehat{Y}_a^{in} \\
\widehat{P}_{Y_b}^{out} \sim  \widehat{P}_{Y_b}^{in}  -  {{G \eta}\over\hbar} \widehat{X}_a^{in}~.
\end{array}
\right.
\label{eqcomments10}
\end{equation}

Choosing the probe observables,
\be\label{eq75}
\widehat{M}=\left({\widehat{X}_{b}\over G}, {\widehat{Y}_{b}\over G}\right)~,
\ee
one finally obtains the noise and disturbance operator,
\be\label{eq76}
\widehat{K}=\left({\widehat{X}_b^{in}\over G} + {{\theta G} \over 2 \hbar} \widehat{P}_{Y_b}^{in}, {\widehat{Y}_b^{in}\over G} - {{\theta G} \over 2 \hbar} \widehat{P}_{X_b}^{in}, -G \widehat{P}_{X_b}^{in}  - {{\eta G^2}\over 2 \hbar} \widehat{Y}_a^{in}, - G \widehat{P}_{Y_b}^{in}  + \frac{\eta G^2}{2 \hbar} \widehat{X}_a^{in}\right)~.
\ee
As anticipated at the beginning of this section, one has additional disturbance terms which are a manifestation of noncommutativity. Notice, in particular, that $\widehat{Y}_a$ is disturbed, unlike the "commutative" case:
\ba
\widehat{D} (\widehat{P}_{X_a}) = - G \widehat{P}_{X_b}^{in} - \frac{\eta G^2}{2 \hbar} \widehat{Y}_a^{in} \label{eq76.1}\\
\widehat{D} (\widehat{P}_{Y_a}) = - G \widehat{P}_{Y_b}^{in} + \frac{\eta G^2}{2 \hbar} \widehat{X}_a^{in} \label{eq76.2}\\
\widehat{D} (\widehat{Y}_a) =  - \frac{\theta G}{\hbar}
\widehat{P}_{X_b}^{in} \label{eq76.3}
 \ea 
To summarize, the effect of noncommutativity can be detected through extra terms in the noise and disturbance due to the new commutation relations, as well as by a disturbance on observables which were undisturbed without noncommutativity in configuration and momentum spaces.

In what follows, we show that Ozawa's measurement-disturbance relation also acquires extra terms due to these new commutation relations.

Turning back to the OUP, Eq.~(\ref{eq5}), and the relations from Eq.~(\ref{eq2}), one typically has for the commutative case
\be\label{eq76} 
||\widehat{N}(\widehat{X}_i)||\propto {1\over G}\hspace{0,2cm},\hspace{0,2cm} ||\widehat{D}(\widehat{P}_i)||\propto G~, 
\ee 
where $i=\widehat{X},\widehat{Y}$ for a $2$-dimensional phase space. Eq.~(\ref{eq2}) then reads
\be\label{eq77} 
\epsilon(\widehat{X}_i)\propto {1\over G}\hspace{0,2cm},\hspace{0,2cm} \chi(\widehat{P}_i)\propto G~. 
\ee 
We now
evaluate the noise and the disturbance in terms of the gain
parameter $G$ and show that the NC corrections add extra terms to
the OUP. For the pair $(\widehat{X}_a, \widehat{P}_{X_a})$, 
\ba\label{eq78}
\epsilon(\widehat{X}_a)&=& \langle\left({\widehat{X}_b^{in}\over G} + G{{\theta } \over 2 \hbar} \widehat{P}_{Y_b}^{in}\right)^2\rangle^{1/2}\nonumber\\
&=&\left(\langle{\widehat{X}_b^{in^2}\over G^2}\rangle + \langle{2{\theta} \over  \hbar} \left\{\widehat{X}_b^{in},\widehat{P}_{Y_b}^{in} \right\}\rangle +\langle G^2{{\theta^2} \over 4 \hbar^2} \widehat{P}_{Y_b}^{in^2}\rangle\right)^{1/2}\nonumber\\
&=&  {{\langle\widehat{X}_b^{in^2}\rangle}^{1/2}\over G} \left(1 + 2G^2{{\theta} \over  \hbar} {\langle \left\{\widehat{X}_b^{in} , \widehat{P}_{Y_b}^{in} \right\} \rangle\over {\langle\widehat{X}_b^{in^2}\rangle}} +G^4{{\theta^2} \over 4 \hbar^2} {\langle\widehat{P}_{Y_b}^{in^2}\rangle\over {\langle\widehat{X}_b^{in^2}\rangle}}\right)^{1/2}~.
\ea
Here $\left\{\widehat{A},\widehat{B} \right\} = \frac{1}{2} (\widehat{A}\widehat{B} + \widehat{B} \widehat{A})$ denotes the anti-commutator. Let us call $\epsilon_C(\widehat{X}_a)={\langle\widehat{X}_b^{in^2}\rangle \over G}$ the "commutative" part of the noise operator, and $k_1= 2 {\langle \left\{\widehat{X}_b^{in} , \widehat{P}_{Y_b}^{in} \right\}\rangle\over {\langle\widehat{X}_b^{in^2}\rangle}}$.
%Since $G<<1$, one should neglect second order terms in $G$.
Thus,
\be\label{eq79}
\epsilon_{NC}(\widehat{X}_a)= \epsilon_C(\widehat{X}_a)\left( 1+ k_1 {\theta\over 2\hbar} G^2\right)+O(\theta^2)~.
\ee
%One points out that this result should be expected as the noise is a classical observable and then should not be affected by the noncommutativity of space-time.
Clearly, the noise has a noncommutative correction due to the noncommutativity between the configuration variables.
The disturbance can be evaluated using the same strategy,
\ba\label{eq80}
\chi(\widehat{P}_{X_a})&=& \langle\left(-G \widehat{P}_{X_b}^{in}  - G^2{{\eta}\over 2 \hbar} \widehat{Y}_a^{in}\right)^2\rangle^{1/2}\nonumber\\
&=&\left(\langle{G^2 \widehat{P}_{X_b}^{in^2}}\rangle + \langle{{\eta} \over  \hbar} G^3\widehat{P}_{X_b}^{in}\widehat{Y}_{a}^{in}\rangle +\langle G^4{{\eta^2} \over 4 \hbar^2} \widehat{Y}_{a}^{in^2}\rangle\right)^{1/2}\nonumber\\
&=& G {\langle{ \widehat{P}_{X_b}^{in^2}}\rangle}^{1/2}  \left(1 + G {{\eta} \over  \hbar} {\langle\widehat{P}_{X_b}^{in}\widehat{Y}_{a}^{in}\rangle\over {\langle{ \widehat{P}_{X_b}^{in^2}}\rangle}} +G^4{{\eta^2} \over 4 \hbar^2} {\langle\widehat{Y}_{a}^{in^2}\rangle\over{\langle{ \widehat{P}_{X_b}^{in^2}\rangle}}}\right)^{1/2}~.
\ea
Thus, defining the ``commutative" part of disturbance as $\chi_C(\widehat{P}_{X_a})=G {\langle{ \widehat{P}_{X_b}^{in^2}}\rangle}^{1/2}$ and  $k_2={\langle\widehat{P}_{X_b}^{in}\widehat{Y}_{a}^{in}\rangle\over {\langle{ \widehat{P}_{X_b}^{in^2}\rangle}}} $, then,
\be\label{eq81}
\chi_{NC}(\widehat{P}_{X_a})= \chi_C(\widehat{P}_{X_a}) \left( 1+ k_2 {\eta\over 2\hbar} G\right)+O(\eta^2)~.
\ee
Notice that
\ba
\left[\widehat{N} (\widehat{X}_a), \widehat{P}_{X_a}^{in} \right] =\left[{{\widehat{X}_b^{in}}\over G} + {{G \theta}\over{2 \hbar}} \widehat{P}_{Y_b}^{in}, \widehat{P}_{X_a}^{in} \right] =0 \label{eq81.1}\\
\left[\widehat{X}_a^{in},\widehat{D} ( \widehat{P}_{X_a}) \right] =\left[\widehat{X}_a^{in}, -G \widehat{P}_{X_b}^{in} - {{G^2 \eta}\over{2 \hbar}} \widehat{Y}_a^{in} \right] = - {{i \theta \eta G^2}\over{2 \hbar}} \label{eq81.2}
\ea
So, unlike the ``commutative" case, the BAE interaction is no longer an independent intervention. However, as before we shall neglect terms of order $O( \eta \theta / \hbar)$.

Of course, the approximations, Eqs. (\ref{eq79}) and (\ref{eq81}), only make sense, provided $k_1$ and $k_2$ are such that
\ba
1+ {{k_1 \theta G^2}\over{2 \hbar}} >0 \label{eqapprox1}\\
1+ {{k_2 \eta G}\over{2 \hbar}} >0 \label{eqapprox2}
\ea
Substituting into Eq.(\ref{eq3}), one finally obtains to lowest order in  $\theta$ and $\eta$
\be\label{eq82}
\epsilon_C(\widehat{X}_a)\chi_C(\widehat{P}_{X_a}) \left( 1+ {{k_1 \theta G^2}\over{2\hbar}}  +{{k_2 \eta G}\over{2 \hbar}} \right) \ge {{\hbar}\over 2}.
\ee

This result suggests that a bound for the NC parameters can be found as a shift with respect to Ozawa's result. It is clear here, that the noncommutative corrections to OUP are associated with the
coefficients $k_1$ and $k_2$. Notice that, it is always possible to choose states  for which $k_2=0$ (which automatically satisfies condition (\ref{eqapprox2})). This is a consequence of the fact that the
interaction is described in the Heisenberg representation. The
initial state is the product state 
\be\label{eq83}
\Psi=\psi\otimes\xi~, 
\ee
where $\psi$ and $\xi$ are the states describing the object and the probe, respectively. Then, one concludes that 
\be\label{eq84}
\langle\Psi~|\widehat{Z}_{a,\alpha}^{in}\widehat{Z}_{b,\beta}^{in}|~\Psi\rangle=\langle\psi~|\widehat{Z}_{a,\alpha}^{in}|~\psi\rangle\langle\xi~|\widehat{Z}_{b,\beta}^{in}|~\xi\rangle~,
\ee 
for every $\alpha=1,..., 4$ and $\beta=1,..., 4$. So, through a translation, it is possible to find a probe state $\xi$, such that 
\be\label{eq85}
\langle\xi~|\widehat{Z}_{b,\beta}^{in}|~\xi\rangle~ =0, 
\ee
for every $\beta=1,..., 4$. This entails that $k_2=0$. Then, OUP becomes 
\be\label{eq86} 
\epsilon_C(\widehat{X}_a)\chi_C(\widehat{P}_{X_a}) \left( 1+{{k_1 \theta G^2}\over {2\hbar}}  \right) \ge {\hbar\over 2}. 
\ee
In this reduced form, all the elements depend only on the probe's
state. It is now manifest that there are probe states for which
the OUP is violated for the BAE model, whereas the noncommutative
version is not. Indeed, choose any state $\xi$ for the probe with
covariance matrix elements $\langle(\widehat{X}_b^{in})^2\rangle$,
$\langle(\widehat{P}_{X_b}^{in})^2\rangle$ and $\langle
\widehat{X}_b^{in} \widehat{P}_{Y_b}^{in}\rangle$ such that 
\be 0
< {\hbar\over 2} \left(1-{{k_1 \theta G^2}\over{2 \hbar}} \right)\le~ \langle(\widehat{X}_b^{in})^2\rangle^{1/2}\langle(\widehat{P}_{X_b}^{in})^2\rangle^{1/2} ~ < {\hbar\over2}.
\label{eq86.1} \ee 
Under these circumstances Eq. (\ref{eq86}) holds to
first order in $\theta$, while the OUP is violated:
\be\label{eq86.2} 
\epsilon_C(\widehat{X}_a)\chi_C(\widehat{P}_{X_a}) < {\hbar\over2}.
\ee

4. {\it Noiseless quadrature transducers}: With the same system and probe previously introduced, suppose now that one has now a measurement interaction as follows. Let $0< T_1 < T_2$, where $T_2$ is the total duration of the measurement interaction. During the time interval $\left[0, T_1 \right]$ the interaction is generated by the Hamiltonian operator
\begin{equation}
\widehat{H}_1 = {1\over T_1} \left(\widehat{P}_{X_b}^{in} \widehat{X}_a^{in} +\widehat{P}_{Y_b}^{in} \widehat{Y}_a^{in} \right)~.
\label{eqnoiseless1}
\end{equation}
This is the same Hamiltonian as for the BAE interaction with $\alpha = T_1^{-1}$. In view of this fact, at time $T_1$ one has:
\begin{equation}
\left\{
\begin{array}{l}
\widehat{X}_a (T_1) = \widehat{X}_a^{in}\\
\widehat{Y}_a (T_1) = \widehat{Y}_a^{in}\\
\widehat{X}_b (T_1) = \widehat{X}_b^{in} +  \widehat{X}_a^{in} \\
\widehat{Y}_b (T_1) = \widehat{Y}_b^{in} + \widehat{Y}_a^{in} \\
\widehat{P}_{X_a} (T_1) = \widehat{P}_{X_a}^{in} - \widehat{P}_{X_b}^{in} \\
\widehat{P}_{Y_a} (T_1) = \widehat{P}_{Y_a}^{in} -  \widehat{P}_{Y_b}^{in} \\
\widehat{P}_{X_b} (T_1) = \widehat{P}_{X_b}^{in}  \\
\widehat{P}_{Y_b} (T_1) = \widehat{P}_{Y_b}^{in}~.
\end{array}
\right.
\label{eqnoiseless2}
\end{equation}
During the subsequent time interval $\left[T_1,T_2 \right]$, the unitary transformation is governed by the Hamiltonian
\begin{equation}
\widehat{H}_2 = -{1\over T} \left(\widehat{P}_{X_a}^{in} \widehat{X}_b^{in} +\widehat{P}_{Y_a}^{in} \widehat{Y}_b^{in} \right).
\label{eqnoiseless3}
\end{equation}
The solution for observable $\widehat{Z} (t)$ during the time interval $\left[T_1,T_2 \right]$ is given by the series:
\begin{equation}
\widehat{Z} (t)= \widehat{Z} (T_1 )+ {{(t-T_1)}\over i \hbar} \left[ \widehat{Z} (T_1), \widehat{H}_2 \right] + {1\over 2!} \left(  {{t-T_1}\over i \hbar}\right)^2 \left[\left[ \widehat{Z} (T_1), \widehat{H}_2 \right] , \widehat{H}_2 \right]+ \cdots~.
\label{eqnoiseless4}
\end{equation}
A straightforward inspection reveals that only the terms up to order $(t-T_1)$ survive for all observables and thus, one gets:
\begin{equation}
\left\{
\begin{array}{l}
\widehat{X}_a (t) = \widehat{X}_a^{in} - {{(t-T_1)}\over T} \widehat{X}_b^{in}\\
\widehat{Y}_a (t) = \widehat{Y}_a^{in}- {{(t-T_1)}\over T}\widehat{Y}_b^{in}\\
\widehat{X}_b (t) = \widehat{X}_b^{in} +  \widehat{X}_a^{in} - {{(t-T_1)}\over T} \widehat{X}_b^{in}\\
\widehat{Y}_b (t) = \widehat{Y}_b^{in} + \widehat{Y}_a^{in} -{{(t-T_1)}\over T} \widehat{Y}_b^{in} \\
\widehat{P}_{X_a} (t) = \widehat{P}_{X_a}^{in} - \widehat{P}_{X_b}^{in} -{{(t-T_1)}\over T} \widehat{P}_{X_a}^{in}\\
\widehat{P}_{Y_a} (t) = \widehat{P}_{Y_a}^{in} -  \widehat{P}_{Y_b}^{in} - {{(t-T_1)}\over T} \widehat{P}_{Y_a}^{in} \\
\widehat{P}_{X_b} (t) = \widehat{P}_{X_b}^{in} +{{(t-T_1)}\over T} \widehat{P}_{X_a}^{in} \\
\widehat{P}_{Y_b} (t) = \widehat{P}_{Y_b}^{in} +{{(t-T_1)}\over T} \widehat{P}_{Y_a}^{in}~,
\end{array}
\right.
\label{eqnoiseless5}
\end{equation}
where $T=T_2-T_1$. Setting $\widehat{X}_a (T_2)= \widehat{X}_a^{out} ,\widehat{Y}_a (T_2)= \widehat{Y}_a^{out}$, etc, we obtain:
\begin{equation}
\left\{
\begin{array}{l}
\widehat{X}_a^{out} = \widehat{X}_a^{in} - \widehat{X}_b^{in}\\
\widehat{Y}_a^{out} = \widehat{Y}_a^{in}- \widehat{Y}_b^{in}\\
\widehat{X}_b^{out} = \widehat{X}_a^{in} \\
\widehat{Y}_b^{out} = \widehat{Y}_a^{in}  \\
\widehat{P}_{X_a}^{out} =  - \widehat{P}_{X_b}^{in} \\
\widehat{P}_{Y_a}^{out} =  -  \widehat{P}_{Y_b}^{in} \\
\widehat{P}_{X_b}^{out} = \widehat{P}_{X_b}^{in} +\widehat{P}_{X_a}^{in} \\
\widehat{P}_{Y_b}^{out} = \widehat{P}_{Y_b}^{in} +  \widehat{P}_{Y_a}^{in}~.
\end{array}
\right.
\label{eqnoiseless6}
\end{equation}

We next resort to the noncommutative algebra, Eqs. (\ref{eq23.1})-(\ref{eq23.4}). At time $t=T_1$ one has (cf.(\ref{eqcomments9}) with $G=1$):
\begin{equation}
\left\{
\begin{array}{l}
\widehat{X}_a^{out} \sim \widehat{X}_a^{in} + {{\theta}\over\hbar} \widehat{P}_{Y_b}^{in}\\
\widehat{Y}_a^{out} \sim \widehat{Y}_a^{in}  - {{\theta}\over\hbar} \widehat{P}_{X_b}^{in} \\
\widehat{X}_b^{out} \sim \widehat{X}_b^{in}+  \widehat{X}_a^{in}  + {{\theta}\over 2 \hbar} \widehat{P}_{Y_b}^{in} \\
\widehat{Y}_b^{out}\sim \widehat{Y}_b^{in}+  \widehat{Y}_a^{in}  - {{\theta}\over 2 \hbar} \widehat{P}_{X_b}^{in} \\
\widehat{P}_{X_a}^{out} \sim \widehat{P}_{X_a}^{in}-   \widehat{P}_{X_b}^{in}  - {{\eta }\over2 \hbar} \widehat{Y}_a^{in}\\
\widehat{P}_{Y_a}^{out} \sim \widehat{P}_{Y_a}^{in}-  \widehat{P}_{Y_b}^{in}  + {{\eta }\over2 \hbar} \widehat{X}_a^{in} \\
\widehat{P}_{X_b}^{out} \sim  \widehat{P}_{X_b}^{in} +  {{ \eta}\over\hbar} \widehat{Y}_a^{in} \\
\widehat{P}_{Y_b}^{out} \sim  \widehat{P}_{Y_b}^{in}  -  {{ \eta}\over\hbar} \widehat{X}_a^{in}~.
\end{array}
\right.
\label{eqnoiseless7b}
\end{equation}

Using the series above, Eq. (\ref{eqnoiseless4}), considering only terms up to second order in $(t-T_1)$ \footnote{Notice that terms beyond $(t-T_1)^2$, will be neglected as they are proportional to ${{\theta\eta}\over\hbar}$ and the noncommutative parameters are presumably small.}, setting $\widehat{X}_a (T_2)= \widehat{X}_a^{out} ,\widehat{Y}_a (T_2)= \widehat{Y}_a^{out}$, etc., and considering again that the noncommutative parameters are small ${\sqrt{\theta\eta}\over\hbar}<<1$, one finally obtains:

\begin{equation}
\left\{
\begin{array}{l}
\widehat{X}_a^{out} = \widehat{X}_a^{in} - \widehat{X}_b^{in}+ {\theta\over\hbar}\left(\widehat{P}_{Y_b}^{in} + {3\over2} \widehat{P}_{Y_a}^{in}\right)\\
\widehat{Y}_a^{out} = \widehat{Y}_a^{in}- \widehat{Y}_b^{in}-{\theta\over\hbar}\left(\widehat{P}_{X_b}^{in} + {3\over2} \widehat{P}_{X_a}^{in}\right)\\
\widehat{X}_b^{out} = \widehat{X}_a^{in} +{\theta\over2\hbar} \widehat{P}_{Y_b}^{in}\\
\widehat{Y}_b^{out} = \widehat{Y}_a^{in}- {\theta\over2\hbar} \widehat{P}_{X_b}^{in} \\
\widehat{P}_{X_a}^{out} =  - \widehat{P}_{X_b}^{in}-{\eta\over2\hbar} \widehat{Y}_{a}^{in} \\
\widehat{P}_{Y_a}^{out} =  -  \widehat{P}_{Y_b}^{in} +{\eta\over2\hbar} \widehat{X}_{a}^{in} \\
\widehat{P}_{X_b}^{out} = \widehat{P}_{X_b}^{in} +\widehat{P}_{X_a}^{in}+ {\eta\over\hbar}\left(\widehat{Y}_{a}^{in} - {3\over2} \widehat{Y}_{b}^{in}\right)\\
\widehat{P}_{Y_b}^{out} = \widehat{P}_{Y_b}^{in} +  \widehat{P}_{Y_a}^{in}- {\eta\over\hbar}\left(\widehat{X}_{a}^{in} - {3\over2} \widehat{X}_{b}^{in}\right)~.
\end{array}
\right.
\label{eqnoiseless9}
\end{equation}

So, one concludes that even for in the noiseless case one has
noncommutative corrections. Moreover, as we can see, the
noncommutativity introduces a ``noise" into the interaction, and
so the transformation is no longer noiseless. If one considers a
probe $\widehat{M}=(\widehat{X}_b,\widehat{Y}_b)$, then the noise
and disturbance are 
\ba\label{eqnoiseless10}
\widehat{K}\!\!=\!\!\left(\!{\theta\over2\hbar}\widehat{P}_{Y_b}^{in},-{\theta\over2\hbar}\widehat{P}_{X_b}^{in},-\widehat{P}_{X_a}^{in}\!\!-\!\!\widehat{P}_{X_b}^{in}\!\!-\!{\eta\over2\hbar}\widehat{Y}_{a}^{in},-\widehat{P}_{Y_a}^{in}\!\! -\!\!\widehat{P}_{Y_b}^{in} \!\!+\!{\eta\over2\hbar}\widehat{X}_{a}^{in}  \right)~. 
\ea Furthermore, in the
configuration variables part, a disturbance also emerges,
\ba\label{eqnoiseless11}
\widehat{D}(\widehat{X_a})=-\widehat{X}_b^{in}+{\theta\over\hbar} ( \widehat{P}_{Y_b}^{in}\!\!+\!\!{3\over2}\widehat{P}_{Y_a}^{in})\nonumber\\
\widehat{D}(\widehat{Y_a})=-\widehat{Y}_b^{in}-{\theta\over\hbar} ( \widehat{P}_{X_b}^{in}\!\!+\!\!{3\over2}\widehat{P}_{X_a}^{in})~.
\ea
As discussed in Ref. \cite{Ozawa}, $\widehat{D}(\widehat{X_a})=0$ is a typical feature of BAE interactions, while $\widehat{N}(\widehat{X}_a)=0$ is a feature of noiseless interactions. Thus, one concludes that the noncommutative extension of a noiseless quadrature transducer transformation is neither noiseless nor BAE.

In terms of OUP, for the noiseless case, the only non-vanishing term is $\sigma(\widehat{X})\chi(\widehat{P}_X)$ as only the disturbance is non-vanishing. In the noncommutative version, the noise becomes
\ba\label{eqnoiseless12}
\epsilon(\widehat{X}_a)&=& \langle\left({\theta\over2\hbar}  \widehat{P}_{Y_b}^{in}\right)^2\rangle^{1/2}\nonumber\\
& =& {\theta\over2\hbar} k_3~,
\ea
where $k_3= \langle\left(\widehat{P}_{Y_b}^{in}\right)^2\rangle$.
The disturbance is
\ba\label{eqnoiseless13}
\chi(\widehat{P}_{X_a})&=& \langle\left(-\widehat{P}_{X_a}^{in}\!\!-\!\! \widehat{P}_{X_b}^{in}\!\!-\!{\eta\over2\hbar} \widehat{Y}_{a}^{in}\right)^2\rangle^{1/2}\nonumber\\
&=&\langle(\widehat{P}_{X_a}^{in}\!\!+\!\! \widehat{P}_{X_b}^{in})^2\rangle^{1/2} \left(1+{\eta\over\hbar}{{\langle \left\{\widehat{P}_{X_a}^{in}\!\!+\!\! \widehat{P}_{X_b}^{in} ,\widehat{Y}_a^{in}\right\}\rangle}\over{\langle(\widehat{P}_{X_a}^{in}\!\!+\!\! \widehat{P}_{X_b}^{in})^2\rangle}}+{\eta^2\over4\hbar^2}{{\langle(\widehat{Y}_a^{in})^2\rangle}
\over{\langle(\widehat{P}_{X_a}^{in}\!\!+\!\! \widehat{P}_{X_b}^{in})^2\rangle}}\right)^{1/2}\nonumber\\
&=&\chi_C(\widehat{P}_X)\left(1+{\eta\over\hbar}k_4+{\eta^2\over4\hbar^2}k_5\right)^{1/2}~,
\ea
where $k_4={{\langle \left\{\widehat{P}_{X_a}^{in}\!\!+\!\! \widehat{P}_{X_b}^{in} ,\widehat{Y}_a^{in}\right\}\rangle}\over{\langle(\widehat{P}_{X_a}^{in}\!\!+\!\! \widehat{P}_{X_b}^{in})^2\rangle}}$ and $k_5={{\langle(\widehat{Y}_a^{in})^2\rangle}
\over{\langle(\widehat{P}_{X_a}^{in}\!\!+\!\! \widehat{P}_{X_b}^{in})^2\rangle}}$. As the noncommutative parameter associated to the momenta, $\eta$, is presumably small \cite{Bertolami}, then
\be\label{eqnoiseless14}
\chi(\widehat{P}_{X_a})=\chi_C(\widehat{P}_{X_a}) \left(1+{\eta\over2\hbar}k_4\right)+O(\eta^2)~.
\ee

Thus, at the lowest non-trivial order of the noncommutative
parameters, the OUP becomes 
\be\label{eqnoiseless15}
{\theta\over2\hbar}k_3 \left[\chi_C(\widehat{P}_{X_a}) +\sigma(\widehat{P}_{X_a})
\right]+\sigma(\widehat{X}_a)\chi_C(\widehat{P}_{X_a})
\left(1+{\eta\over2\hbar}k_4\right)\ge {\hbar\over2}~. 
\ee

As before one can find states that violate the commutative OUP, while satisfying the noncommutative version of the OUP, Eq. (\ref{eqnoiseless15}).

5. {\it Discussion}: In this work, Ozawa's noise-disturbance relation is extended to the framework of phase-space NCQM. We considered first a BAE quadrature amplifier system. This system, which
refers to an independent intervention for a pair of quadrature operators $\widehat{X}_a,\widehat{P}_{X_a}$, ceases to be so once the phase-space noncommutative algebra is considered. Moreover, a second pair of quadrature operators $\widehat{Y}_a,\widehat{P}_{Y_a}$ are found to be disturbed by a measurement of $\widehat{X}_a$, which is in contrast with what happens in ordinary quantum mechanics. We also found that, as expected, extra terms appear in the OUP.

As for noiseless quadratures transducers, noncommutativity introduces a noise term and so the interaction is no longer noiseless. In fact, the noiseless case transforms into a new form of interaction which is neither noiseless, nor a BAE interaction.

Finally, we have shown in both cases, that there are states that violate the OUP, but are in agreement with the NCOUP. This shows that the NCQM encompasses more states than the standard QM. Thus, experimentally, a tiny imprint of noncommutativity could be identified in quantum systems, if an effective deviation from OUP were detected.

\vspace{1,5cm}

{\it Acknowledgements}:
The work of CB is supported by Funda\c{c}\~{a}o para a Ci\^{e}ncia e a Tecnologia (FCT) under the grant SFRH/BPD/62861/2009. The work of AEB is supported by the Brazilian Agency CNPq  under the grant 300809/2013-1. The work of OB is partially supported by the FCT project
PTDC/FIS/111362/2009. N.C. Dias and J.N. Prata have been supported by the FCT grant PTDC/MAT/099880/2008.

%\end{acknowledgements}

\end{document}